\documentclass[prb,showpacs,preprintnumbers,amsmath,amssymb,twocolumn]{revtex4}

\usepackage{epsfig}
\usepackage{graphicx}
\usepackage{dcolumn}
\usepackage{bm}

\newcommand{\vk}{{\bf k}}
\newcommand{\vQ}{{\bf Q}}

\newcommand{\vq}{{{\bf q}}}
\newcommand{\vK}{{{\bf K}}}
\newcommand{\vKp}{{{\bf K}'}}

\newcommand{\vf}{v_{\rm F}}
\newcommand{\va}{{\bf a}}
\newcommand{\vb}{{\bf b}}

\newcommand{\kf}{k_{\rm F}}

\newcommand{\tp}{t_\perp}
\newcommand{\pk}{\phi_{\bf k}}

\newcommand{\mk}{| {\bf k} |}
\newcommand{\ek}{e^{i \pk}}
\newcommand{\emk}{e^{-i \pk}}

\begin{document}

\title{Electron-Phonon Coupling and Raman Spectroscopy in Graphene}
\author{A. H. Castro Neto}
\affiliation{Department of Physics, Boston University, 590 
Commonwealth Avenue, Boston, MA 02215,USA}
\author{Francisco Guinea}
\affiliation{Instituto de Ciencia de Materiales de Madrid, CSIC,
 Cantoblanco E28049 Madrid, Spain}

\begin{abstract}
We show that the electron-phonon coupling in graphene, in contrast with 
the non-relativistic two-dimensional electron gas, leads to shifts in 
the phonon frequencies that are non-trivial functions of the electronic 
density. These shifts can be measured directly in Raman spectroscopy. 
We show that depending whether the 
chemical potential is smaller (larger) than half of the phonon frequency, 
the frequency shift can negative (positive) relative to the neutral 
case (when the chemical potential is at the Dirac point), respectively. 
We show that the  use of the static response function to calculate these 
shifts is incorrect  and leads always to phonon softening.  In samples with 
many layers, we find 
a shift proportional to the carrier concentration, and a splitting of the 
phonon frequencies if the charge is not homogeneously distributed. 
We also discuss the effects of edges in the problem.
\end{abstract}


\maketitle

\section{Introduction}
\label{intro}

The discovery of graphene, a thermodynamically stable two-dimensional (2D) 
crystal \cite{Netal04}, whose electronic properties, described in terms of 
a half-filled $\pi$-electronic band with Dirac electrons, can be controlled 
externally, has stirred great interest in the scientific community since the 
demonstration of a theoretically predicted \cite{nuno2006_long,GS05} 
anomalous integer quantum Hall effect \cite{Novoselov2005,Zhang2005}. Unlike
other 2D electronic systems, such as MOSFET heterostructures, graphene is 
easily accessible to optical probes. Furthermore, in contrast to ordinary 
semiconductors where the different types of disorder can be distinguished 
through the temperature dependence of the transport properties
\cite{shockley}, graphene does not show any strong temperature or magnetic 
field dependence in its electronic transport \cite{morozov_prl_06} that
allows an easy discrimination between different types of impurities. Hence, 
local probes such as scanning tunneling microscopy (STM) and single electron
transistor probes, will play a fundamental role in the understanding of
nature the effects of disorder in graphene-based systems. 

Raman spectroscopy has been one of the most successful experimental 
methods used to study these systems 
\cite{ferrari_06,eklund_06,ensslin_06,pinczuk}.
In particular, it has been shown that it is possible to measure the number 
of graphene layers on a SiO$_2$ substrate with great accuracy, leading to an 
efficient and fast method to characterize graphene {\it in situ}. Another 
interesting feature of these measurements is that, even for a single graphene 
layer, the phonon frequency measured in Raman shifts by a few wavenumbers, 
from point to point in space \cite{ferrari_06,eklund_06,ensslin_06}. Moreover, 
the observation of a D-line, which is Raman forbidden in translational
invariant graphene, indicates the presence of disorder in the samples at 
electronic scale. 

We show that this Raman shift can be associated with the earlier experimental 
evidence for charge inhomogeneity in undoped, unbiased, graphene \cite{thanks_geim,morozov_prl_06}. Therefore, Raman spectroscopy can be used to {\it map}
the disorder in graphene layers, and hence, help to shed light on the nature 
of the disorder scattering in these materials. The understanding of the
nature of impurity scattering in graphene is fundamental not only for the 
development of electronic devices based on carbon, but also may help to solve 
theoretical puzzles such as the discrepancy found between the theoretically 
predicted {\it universal} value of the conductivity \cite{F86}, $4 e^2/(\pi h)$, and its experimentally observed \cite{Novoselov2005} value of $4 e^2/h$  
(the so-called ``mystery of the missing $\pi$''), and the absence of 
weak-localization effects \cite{morozov_prl_06} (a topic that has generated 
intense theoretical debate \cite{mccann_altshuler_06_short,morpurgo_guinea,dima_06,ziegler_06,kentaro_06,aleiner_efetov_06,altland_06}). 

In this paper we show that the shift in the phonon frequency in graphene has
its origin on the polarization of the electrons due to the ion motion. 
Since graphene is a perfect hybrid between a metal and a semiconductor
there are two contributions to the polarization function: one comes from
intra-band transitions and another that originates on inter-band transitions.
We show that the simplest approximation based on the static response is incorrect and
predicts a reduction of the phonon frequency (softening of the lattice). The
correct dynamic response is used to calculate the phonon frequency shift 
and it is shown that the phonon frequency can either decrease (softening) or  
increase (hardening) depending on whether the phonon frequency is either 
larger or smaller than twice the chemical potential, respectively. We also 
show that the intra-band dynamic response {\it vanishes} at long wavelengths 
in a 
translationally invariant graphene sheet, while the inter-band contribution
is finite. Nevertheless, in disordered graphene we expect the intra-band
contribution to be of the order of the inter-band one, indicating that disorder 
is important for the measurement of the Raman shift in graphene. 

The paper is organized as follows: in section \ref{model} we present the
model for the electrons, phonons, and their coupling in graphene; section
\ref{single} discusses the problem of the shift of the phonon frequency due
to the electronic polarization in graphene and we consider both the static
and the dynamic response; in section \ref{bilayer} we examine the problem of
phonon frequency shifts in bilayers and multilayers within the same
framework; section \ref{conclusions} contains a discussion of the problem of
edges in finite samples and also the main conclusions of our work. We have
also included one appendix with the details of an analytical model for the
in-plane phonon modes in graphene, and also discuss the effect of defects and
edges in the phonon spectra.

\section{The Model.}
\label{model}

In the absence of disorder the Hamiltonian for electrons and phonons in 
graphene can be written as ${\cal H} ={\cal H}_E+{\cal H}_P+{\cal H}_{E-P}$, 
where (we use units such that $\hbar = 1 = k_B$),
\begin{eqnarray}
{\cal H}_E = -t_0 \sum_{\langle i,j \rangle} \left(c^{\dag}_{A,i} c_{B,j} + {\rm h.c.}\right) - \mu 
\sum_{i,a=A,B} c^{\dag}_{a,i} c_{a,i} \, ,
\end{eqnarray}
is the free electron Hamiltonian, where $\mu$ is the chemical potential, 
$c_{a,i}$ ($c^{\dag}_{a,i}$) annihilates
(creates) and electron on sublattice $a = A,B$ on site ${\bf R}_i$ in the honeycomb lattice 
(spin indices are omitted throughout the paper), 
and $t_0 \approx 2.7$ eV, is the nearest neighbor hopping energy.  $H_P$ is the phonon Hamiltonian:
\begin{equation}
{\cal H}_P = \sum_{\vq , i} \omega_{\vq i} \, b^\dag_{\vq i} b_{\vq i} \, ,
\end{equation}
where $b_{{\bf q},a}$ ($b^{\dag}_{{\bf q},a}$) annihilates (creates) a phonon with
momentum ${\bf q}$, and $i =$ TA , LA , TO , L0, are the four phonon modes \cite{wirtz_04}. 
In the following, we focus on the transverse optical (TO) modes
near the $\Gamma$ and $K$ and $K'$ points of the Brillouin zone. The TO band
shows little dispersion with a frequency $\omega_0 \approx 0.19$ eV.

We assume that the electron-phonon coupling arises from the modulation by the
phonons of the Carbon-Carbon distance, $a=1.42$ \AA, which leads to a change in
the nearest neighbor hopping $t_0$. The dependence of $t_0$ on distance $l$ has
been extensively studied \cite{DSM77,BCP88}:
\begin{equation}
\partial t_0/\partial l = \alpha \approx 6.4 \, {\rm eV \AA}^{-1} \, .
\end{equation}
The resulting electron-phonon interaction is:
\begin{widetext}
\begin{eqnarray}
{\cal H}_{E-P} &=  & (\partial t_0/\partial l) \sum_{\vk , \vq}
 c^\dag_{A {\vk}} c_{B {\vk + \vq}}    
\left\{  
x_{A \vq} \left[ 1 - e^{i ( \vk + \vq ) {\bf a}}/2 -
e^{i ( {\bf k} + \vq ) \cdot {\bf b}}/2 \right] 
- x_{B {\bf q}} 
\left[ 1 - 
e^{i  {\bf k}  \cdot {\bf a}}/2 -
e^{i  {\bf k} \cdot {\bf b}}/2 \right] 
+ \right.
\nonumber \\  
&+ &\left.
(\sqrt{3}/2) y_{A \vq} \left[
e^{i ( \vk + \vq ) \cdot \va} - e^{i (\vk + \vq ) \cdot \vb} \right] -
(\sqrt{3}/2) y_{B \vq} \left[
e^{i  \vk \cdot \va} - e^{i \vk \cdot \vb} 
 \right] 
 \right\} 
 + {\rm h.c.}
\label{ep}
\end{eqnarray}
\end{widetext}
 where $\va$ and $\vb$ are the unit vectors of the honeycomb lattice,
and $x_{a \vq} , y_{a \vq}$ ($a=A,B$) are given by the polarization of the phonon
of wavevector $\vq$. They can be written as:
\begin{equation}
\left( \begin{array}{c} x_{A \vq} \\ y_{A \vq} \\ x_{B \vq} \\ y_{B \vq}
  \end{array} \right) \equiv \frac{1}{\sqrt{2 M_C \omega_{\vq}}} 
\left( b^\dag_{\vq} + b_{- \vq} \right) \left(
  \begin{array}{c} \alpha_1 \\ \alpha_2 \\ \alpha_3 \\ \alpha_4 \end{array}
\right) \, ,
\end{equation}
where $M_C = 1.2 \times 10^{4} m_e$ is
the carbon mass ($m_e$ is the electron mass), and the vector $( \alpha_1 ,
\alpha_2 , \alpha_3 , \alpha_4 )$ is normalized to one.

In order to obtain the polarizability of the TO mode, we use a central force 
model (see appendix \ref{anamodel}) which leads to a phonon dispersion which 
can be calculated analytically \cite{G81}. This model is adapted from similar 
models for tetrahedrally bonded lattices \cite{WA72}. The details of the
model are described in the appendix \ref{anamodel}, where it is illustrated
by some applications. The honeycomb lattice, even in the limit when the bonds 
are incompressible, can have shear deformations, leading to a vanishing shear 
modulus. Because of it, the model shows a flat transverse acoustical branch 
at zero energy. The optical modes, on the other hand, induce significant 
changes in the bond lengths. We focus here on a single optical mode, whose 
energy we take from experiments. The polarization is fixed by symmetry 
considerations. Hence, the model is needed only to describe the coupling to 
the electrons. The only coupling consistent with nearest neighbor tight 
binding model used to describe the $\pi$ bands is the one that we are using. 

The polarization of the non-degenerate mode at the K point in
the Brillouin zone is:
\begin{equation}
( \alpha_1 , \alpha_2 , \alpha_3 , \alpha_4 ) = \left(1/2 ,i/2,- 1/2 ,i/2
\right) \, ,
\label{polarization}
\end{equation}
and we have a doubly degenerate mode (a Dirac phonon) with polarizations:
\begin{eqnarray}
( \alpha_1 , \alpha_2 , \alpha_3 , \alpha_4 ) &= &\left(1/\sqrt{2} ,
  -i/\sqrt{2} , 0 , 0 \right) \, , \nonumber \\
( \alpha_1 , \alpha_2 , \alpha_3 , \alpha_4 ) &= &\left( 0 , 0 , 1/\sqrt{2}
  ,i/\sqrt{2}  \right) \, .
\label{polarization_2}
\end{eqnarray}
For comparison, the polarization of the two optical modes at the $\Gamma$
point can be written as:
\begin{eqnarray}
( \alpha_1 , \alpha_2 , \alpha_3 , \alpha_4 ) &= &\left(1/\sqrt{2} ,
  0 , -1/\sqrt{2} , 0 \right) \, , \nonumber \\
( \alpha_1 , \alpha_2 , \alpha_3 , \alpha_4 ) &= &\left( 0 , 1/\sqrt{2} , 0
  , -1/\sqrt{2} \right) \, .
\label{polarization_3}
\end{eqnarray}

\section{Single layer graphene}
\label{single}

We are interested in the modification induced by electronic transitions 
of the frequency of a phonon with wavevector $\vQ$. The electronic
transitions which describe these processes are given, approximately, by:
\begin{eqnarray}
{\cal H}_{\vQ} &\equiv & 3 \alpha/2 \sum_{\vk} c^\dag_{A \vK} c_{B \vKp + \vk} \left( x_{A \vQ} -
  x_{B \vQ} + \right. \nonumber \\ &+ &\left. i y_{A \vQ} + i y_{B \vQ}
  \right) + {\rm h.c.} \, ,
\label{ep_2}
\end{eqnarray}
where we assume that the main contribution arises from transitions close to
the Fermi level. In this limit, we can use the continuum limit and expand the 
energy of the electrons around the $K$ and $K'$ points, leading
to the Hamiltonian:
\begin{eqnarray}
{\cal H}_0 = \left(
\begin{array}{cccc} -\mu & \vf \mk \ek &0 &0 \\ \vf \mk \emk &-\mu &0 &0 \\ 0
  &0 &-\mu & \vf \mk \emk \\ 0 &0 
&\vf \mk \ek &-\mu \end{array} \right) \, ,
\label{h0}
\end{eqnarray}
where $\vf = 3 t_0 a/2 \approx 6$ eV \AA \, is the Fermi-Dirac velocity, and
$\phi_{{\bf kß}} = \arctan(k_y/k_x)$ 
is the angle in momentum space.

\begin{figure}
\begin{center}
\includegraphics*[width=5cm]{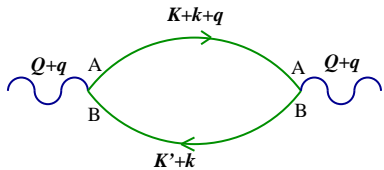}
\end{center}
\caption{Diagram which describes the modification of the phonon propagator
  due to electron-hole transitions. See text for details. }
\label{phonon_ren}
\end{figure} 

A typical diagram which describes the renormalization of the phonon
propagator in second order perturbation theory  is given in 
Fig.~[\ref{phonon_ren}]. The convolution of electronic Green's functions 
shown in the diagram is formally identical to the charge susceptibility of 
graphene:
\begin{eqnarray}
\chi ( \vq , \omega ) &= &\int d^2 \vk d \omega' \left[ G_{AA}^{{\rm occ}} ( \vk ,
  \omega' ) G_{BB}^{{\rm empty}} ( \vk + \vq , \omega + \omega' ) +
  \right. \nonumber \\ &+ &\left. G_{AB}^{{\rm occ}} (
  \vk , \omega' ) G_{BA}^{{\rm empty}} ( \vk + \vq , \omega + \omega' ) \right] \, ,
\label{susc}
\end{eqnarray}
where $G_{ab}^{{\rm occ}}({\bf k},\omega)$ ($G_{ab}^{{\rm empty}}({\bf k},\omega)$)
with $a,b = A,B$ is the electronic Green's function for the occupied (empty)
states.  The main difference between (\ref{susc}) and the charge
susceptibility 
of graphene is that the charge susceptibility includes an overlap factor 
which suppresses completely transitions between the valence and conduction 
band at $\vq=0$. 

From Fig.~[\ref{phonon_ren}] we can immediately obtain the shift in the phonon
frequency due to the polarization of the graphene layer due to particle-hole
excitations:
\begin{equation}
\delta \omega_{\vQ} =  \frac{27 \!\sqrt{3} a^2\!
  |\alpha_1-\alpha_2+i\alpha_3+i\alpha_4|^2\!}{16 M_C \omega_{\vQ}} \!\! 
\left( \frac{\partial  t_0}{\partial l} \right)^2 \!\!\! \chi(\omega_\vQ,q \to 0) \, ,
\label{main}
\end{equation}
where we have used that $\Omega = 3 \sqrt{3} a^2 / 2$ is the area of the unit cell.

Graphene, from the electronic point of view,
is a hybrid between a metal and a semiconductor: the polarization involves 
not only inter-band excitations
(as in the case of the ordinary electron gas) but also intra-band excitations
(as in the case of a semiconductor). 
The full susceptibility $\chi$ can
be separated into an intra- and an inter-band contributions:
\begin{equation}
\chi(q,\omega) = \chi^{\rm inter}(q,\omega) + \chi^{\rm intra}(q,\omega) \, .
\end{equation}
The intra-band contribution to the susceptibility was originally calculated 
by K.W.-K. Shung for graphene in
ref.~[\onlinecite{shung}] and more recently it has appeared on 
refs.~[\onlinecite{Gon94,GGV99,A06,sols}].

\subsection{Static approximation}

A commonly used approximation on the electron gas problem is to replace
the dynamical response $\chi(\omega=\omega_Q,q=0)$ by the static one:
$\chi(\omega=0,q=0)$. This approximation is usually justified in ordinary
metals because the Fermi energy $\mu$ is much larger than the phonon
frequency so that the phonons respond to a time averaged electron
distribution. In graphene, however, this is not necessarily so. 
In what follows we will study the effect of an static response and
compare it with what happens when a dynamic response is calculated
instead. We will show that these two approximations give very 
different results.  

At finite doping, the compressibility sum rule for the charge
  susceptibility leads to the equation:
\begin{equation}
\lim_{\vq \rightarrow 0} \chi^{{\rm intra}} ( \vq , \omega = 0 ) = - D ( \mu ) = - \frac{2
  \mu}{\pi \vf^2} \, ,
\label{static}
\end{equation}
where $D ( \mu )$ is the density of states at the Fermi level.
The number of carriers per unit cell, $n_\Omega$, is given by, 
\begin{equation}
n_\Omega = \frac{3 \sqrt{3}}{2 \pi} ( \kf a )^2 \, .
\label{nomega}
\end{equation}
The coupling to optical phonons involves terms which couple the two
sublattices $A$ and $B$, eq.(\ref{ep_2}), while the charge susceptibility
leading to eq.(\ref{static}) is the response to the total 
charge density operator:
$\rho(\vq) = \sum_{\vk} (c^\dag_{A \vk + \vq} c_{A \vk} +
c^\dag_{B \vk + \vq} c_{B \vk})$. This change modifies the overlap factors, 
leading to an angular factor, when $\vq \rightarrow 0$, equal to 
$\sin^2 ( \theta )$ or $\cos^2 (\theta )$, depending on the 
polarization of the phonon. The average of this term gives rise to a factor 
$1/2$ with respect to the (diagonal) charge susceptibility calculated in
refs.~[\onlinecite{shung,Gon94,GGV99,A06,sols}].

In addition, we have:
\begin{equation}
\chi^{\rm inter}(q \to 0,\omega=0) = - \frac{1}{\pi^2}   \int_{\kf}^{\Lambda}  k d k \int_0^{2 \pi}d \theta \frac{
  \cos^2 ( \theta )}{\epsilon_k}  \, ,
\label{inter}
\end{equation}
where the angular factor, as for the intraband susceptibility, depends on the
polarization of the phonon and whose average is always $1/2$.

The main contribution to the integral in eq.(\ref{inter}) comes from $k \sim \Lambda$, where
$\Lambda$ is the high energy cut-off, so that this expression depends on
details of the bands at high energies away from the Dirac point. 
Nevertheless, the change of the susceptibility with electronic density
is independent of the cut-off and can be readily calculated:
\begin{eqnarray}
\lim_{\vq \rightarrow 0} \left(\chi^{\rm inter} ( \mu ) - \chi^{\rm inter} ( 0 ) \right)=
 \nonumber
\\
\delta \chi^{\rm inter}(q \to 0,\omega=0)
\approx - \frac{\mu}{\pi \vf^2} \, .
\label{static_2}
\end{eqnarray}
which is of the same magnitude as the intra-band shift, eq.(\ref{static}).

Inserting (\ref{static}), (\ref{static_2}) and (\ref{nomega}) into
(\ref{main}), we find, in addition to a density independent shift: 
\begin{equation}
\delta \omega_{\vQ} = - \frac{9}{\sqrt{6 \pi}} \left( \frac{\partial
  t_0}{\partial l} \right)^2 \frac{n^{1/2}_\Omega}{M_C
  \omega_{\vQ} t_0}  \, .
\label{shiftstatic}
\end{equation}
Expressing $\delta \omega_{\vQ}$ (in eV), and replacing $n_\Omega$ by the
density per unit area, $n$ (expressed in cm$^{-2}$), we find:
\begin{equation}
\delta \omega_{\vQ} ( {\rm eV} ) \approx - 3 \times 10^{-9} n^{1/2}( {\rm cm^{-2} } )  \, ,
\label{shift_2static}
\end{equation}
is the expression for the shift of the phonon frequency in the static 
approximation.  For typical electron (or hole) densities, 
$n \approx 10^{11}-10^{12} \, {\rm  cm^{-2}}$, the shifts are of the order 
of a few wavelengths (or degrees Kelvin), within experimental accuracy. 

Let us first notice that this result indicates that there is a decrease of
the phonon frequency, that is, a {\it softening} of the lattice. This result
is generically expected on physical grounds since a high density of electrons
leads to the screening of the ion-ion interactions, reducing the elastic
coupling in the lattice, and hence leading to a softening of the phonons. 

\subsection{Dynamic approximation}

The real part of the intra-band susceptibility is given by \cite{sols}:
\begin{eqnarray}
\Re[\chi^{{\rm intra}}(\omega,q \to 0)] \approx - \frac{q^2}{2 \pi \omega} \left\{\frac{2 \mu}{\omega}+
\frac{1}{2} \ln\left|\frac{2 \mu-\omega}{2 \mu+\omega}\right| \right\} \, .
\label{dynamicintra}
\end{eqnarray}
This expression is rather different from the static result (\ref{static}). 
We note that the limits of $\omega=0$ and $q\to 0$ with $\omega \to 0$ and 
$q = 0$ do not commute. Moreover, we clearly see that (\ref{dynamicintra}) changes 
behavior whether $\omega/(2 \mu)$ is smaller or larger than one, and 
the susceptibility has a logarithmic singularity in $\omega =  2 \mu$. 

For $2 \mu \gg \omega$ a self consistent calculation of the polarizability
shows the existence of the two dimensional plasmon, which needs to be taken
into account. We find:
\begin{eqnarray}
\Re[\chi^{{\rm intra}}(\omega \ll 2 \mu,q \to 0)] \approx \frac{- q^2 \mu}{\pi [ \omega^2 -
  \omega_{pl}^2 ( q ) ]} \, ,
\label{large}
\end{eqnarray}
where $\omega_{pl} ( q ) = \sqrt{( 2 e^2 \mu q ) / \epsilon_0}$ is the
plasmon frequency ($e$ is the electric charge and $\epsilon_0$ the dielectric
constant of graphene). For $2 \mu \ll \omega$ we find:
\begin{eqnarray}
\Re[\chi^{{\rm intra}}(\omega \gg 2 \mu,q \to 0)] \approx \frac{4}{3 \pi} \frac{q^2
  \mu^2}{\omega^4} \, .
\label{small}
\end{eqnarray}
Observe the change of sign in the expression of the susceptibility in the two
limits. More importantly, one can clearly see that these expressions vanish 
when $q \to 0$. This effect 
occurs because at $q=0$ the states associated with these transitions 
are orthogonal.  
Hence, in a system with translational invariance the 
intra-band transitions give no contribution. 
Nevertheless, in the presence of disorder (or a finite sample), the 
electron mean free path, $\ell$, (or the system size, $L$) acts naively as a 
infrared cut-off and one would expect to see a non-zero effect.
Replacing (\ref{large}) and (\ref{small}) into (\ref{main}), and assuming
that $\omega_{\vQ} \gg \omega_{pl}$,  we find:
\begin{eqnarray}
\delta \omega^{{\rm intra}}_{\vQ}(2 \mu>\omega_Q) \!\!&\approx&\!\! \frac{-9}{2\sqrt{6 \pi}} 
\!\! \left( \!\frac{\partial t_0}{\partial l} \!\right)^2 
\!\!\!\! \frac{n^{1/2}_\Omega}{M_C \omega_{\vQ} t_0} 
\!\!\left(\!\frac{\vf q}{\omega_{\vQ}}\!\right)^2 \!\!\!,
\\
\delta \omega^{{\rm intra}}_{\vQ}(2 \mu<\omega_Q) \!\!&\approx&\!\! \frac{9 \sqrt{\pi}}{\sqrt{2}} 
\!\! \left( \!\frac{\partial  t_0}{\partial l} \!\right)^2 
\!\!\!\! \frac{n^{3/2}_\Omega}{M_C \omega_{\vQ} t_0}  
\!\! \left(\!\frac{t_0 \vf q}{\omega^2_{\vQ}}\!\right)^2 \!\!\!,
\label{shift}
\end{eqnarray}
Expressing $\delta \omega_{\vQ}$ in eV, and replacing $n_\Omega$ by the
density per unit area, $n$, expressed in cm$^{-2}$, we find:
\begin{eqnarray}
\delta \omega^{{\rm intra}}_{\vQ} (2 \mu > \omega_Q) &\approx&  + 1.4 \times 10^{-6}
n^{1/2} \, (q a)^2 \, ,
\\
\delta \omega^{{\rm intra}}_{\vQ} (2 \mu < \omega_Q) &\approx&  -  1.0 \times 10^{-18} 
n^{3/2} \, (q a)^2 \, .
\label{shift_2}
\end{eqnarray}
We stress, once again, that this shift vanishes as $q \to 0$ and hence
there should be no shift in the phonon frequency in a translationally
invariant graphene sheet. Nevertheless, in the presence of disorder
this is not necessarily the case. 

In order to include disorder in the calculation one would have to dress the 
fermion propagators in Fig.~[\ref{phonon_ren}] by disorder and include vertex 
corrections to that diagram. These calculations are beyond the scope of this 
paper. Instead, we will follow a naive approach and simply introduce a
cut-off in $q$ of the order of the inverse of the electron mean free path, 
$\ell$, which is known to be of order of $0.1 \mu {\rm m}$ in these systems 
\cite{morozov_prl_06}. For typical electron (or hole) densities, 
$n \approx 10^{12} -10^{13} \, {\rm cm^{-2}}$, and $\ell \approx 10^3 a$ 
($\ell \approx 0.1 \mu {\rm m}$)  the shifts are of the order of 
$10^{-6} - 10^{-5}$ eV. For these concentrations and wavelengths, 
the plasmon frequency is $\omega_{pl} \approx 0.01 - 0.04$eV, so that the 
assumptions leading to eq.(\ref{shift}) are justified. 
Notice that while for large doping, $\mu >\omega_Q/2$, the intra-band
contribution leads to a {\it
  hardening} of the phonon, for low doping, $\mu<\omega_Q/2$ there is 
{\it softening} of the phonon mode which depends directly on the amount of 
disorder in the system.

The inter-band susceptibility is:
\begin{equation}
\Re[\chi^{\rm inter} (q \to 0,\omega)] \approx - \frac{1}{\pi}
 {\cal P} \int k dk \frac{4 \epsilon_k}{\omega^2 - 4 \epsilon_k^2} \, .
\end{equation}
As in the static case, we find a large contribution which is independent of
the carrier concentration and depend on the high-energy cut-off. 
As before, we consider only the density dependent contribution:
\begin{eqnarray}
\Re[\delta \chi^{\rm inter} (q \to 0, \omega)] \approx  -
  \frac{\mu}{\pi \vf^2} - \frac{\omega}{4 \pi \vf^2} \log \left| \frac{\omega
  - 2 \mu}{\omega + 2 \mu} \right| \, .
\label{dynamicinter}
\end{eqnarray} 
The first term in this expression reproduces the static limit of $\chi^{{\rm
  inter}}$. 
The second term gives a correction which is more important
  for $\omega \sim 2 \mu$, and cancels the static contribution as 
$\omega_{\vQ} /  \mu \rightarrow \infty$. Comparing (\ref{dynamicintra}) with
(\ref{dynamicinter}) we find that:
\begin{eqnarray}
\Re[\chi^{{\rm intra}}(\omega, q=\omega/\vf)] 
= \Re[\delta \chi^{\rm inter}(\omega,q=0)] \, ,
\end{eqnarray}
implying, from (\ref{shift_2}), that:
\begin{eqnarray}
\delta \omega^{{\rm inter}}_{\vQ} (2 \mu > \omega_Q) &\approx&  + 3 \times 10^{-9}
n^{1/2} \, ,
\\
\delta \omega^{{\rm inter}}_{\vQ} (2 \mu < \omega_Q) &\approx&  -  2 
\times 10^{-21} n^{3/2} \, .
\label{shift_2_inter}
\end{eqnarray}
Once again, for $\mu >\omega_Q/2$, the inter-band
contribution leads to a {\it
  hardening} of the phonon, while for $\mu<\omega_Q/2$ there is 
{\it softening} of the phonon mode. Notice, that the numerical
value of the inter-band contribution is small for densities of
order $10^{12}$ cm$^{-2}$ when compared with the intra-band
contribution estimated in the presence of disorder. It may well
be that in disordered graphene the intra-band transitions dominate
over the inter-band transitions. Hence, the final result may
vary with the amount of disorder in the samples.

\section{Bilayers}
\label{bilayer}

The previous analysis can be extended to a bilayer system. For simplicity, we
consider here the static limit only. The model for 
the in-plane phonons considered before needs no changes. The shift in the
phonon frequency is also given by the electronic susceptibility shown in
Fig.~[\ref{phonon_ren}], and given in eq.(\ref{susc}). 

In a bilayer, however,
the wavefunctions corresponding to the low energy electronic states have a
small amplitude, $a_{A \vk} \sim \vk | {\vk} | / t_\perp$ in the orbitals 
hybridized through the hopping $t_\perp$ with an orbital in the next layer
($t_{\perp} \approx 0.3$ eV is the inter-layer hopping energy). The relevant
susceptibility involves a convolution of the Green's function of sites in
both sublattices, so that the reduction in low energy spectral weight at the
sites with a neighbor in the next layer will reduce the susceptibility.

The amplitude of an electronic wavefunction at energy $\ek \approx \vf^2
  \mk^2 / \tp^2
  \ll 1$ is of order $a_B \sim V^{-1/2}$, where $V$ is the volume of the
  system,  on sites of the sublattice not
  connected to the second layer, defined as sublattice $B$. The amplitude on
  sites in sublattice $A$, where the sites 
are connected to the second layer is of  order $a_A
  \sim \vf \mk / \tp \times a_B$.  Then, the contribution to the
  susceptibility from the low energy electron-hole pairs is of the form:
\begin{equation}
\chi_{{\rm bilayer}}^{\rm intra} ( \vq , \omega = 0 ) \approx \int d \vk
  \frac{\vf^2 \vk ( \vk 
  + \vq )}{t_\perp^2} \frac{n_{\vk + \vq} - n_{\vk}}{\epsilon_{\vk + \vq} -
  \epsilon_{\vk}} \, ,
\label{susc_bilayer}
\end{equation}
The same suppression applies to inter-band transitions, as, in any case, the
modulation of the hopping involves transitions from the $A$ sublattice to the
$B$ sublattice. Using dimensional
arguments, which are also valid for single layer graphene, at the neutrality
point we find, 
$
\chi_{{\rm bilayer}}^{\rm intra} ( \vq , \omega = 0 ) \propto | \vq |^2/t_\perp \, ,
$
and, at finite fillings, 
\begin{equation}
\lim_{| \vq | \rightarrow 0} \chi_{{\rm bilayer}}^{\rm intra} ( \vq , \omega
= 0 ) \propto \kf^2/t_\perp \,  .
\label{shift_bilayer_1}
\end{equation}
As in the single layer case, the inter-band susceptibility includes a
contribution determined by the high energy cut-off, and  density dependent
term, which, also on dimensional grounds, depends on density as the intra-band
susceptibility, (\ref{shift_bilayer_1}).

For the inter-band contribution, using the reduction in the amplitude at the
$A$ sublattice mentioned earlier, we find:
\begin{equation}
\chi_{{\rm bilayer}}^{\rm inter} ( \vq , \omega = 0 ) \propto \frac{\vf^2
  \kf^2}{\tp^2} \int_{\kf}^{\Lambda_{\rm bil}} d \vk
  \frac{\mk}{\epsilon_{\vk}} \, ,
\label{susc_bilayer_2}
\end{equation}
where $\Lambda_{\rm bil} = \tp / \vf$ is a high momentum cut-off above which
the assumption that $\vf k \ll t_\perp$ ceases to be valid. The integral in this
expression has a logarithmic dependence on $\Lambda_{\rm bil}$, similar to
the logarithmic divergences which characterize the charge susceptibility of a
bilayer\cite{Nilsson06a}. 

As in the case of the single layer, when we insert eq.(\ref{susc_bilayer_2})
into the expression for the shift in the phonon frequency, we find an term
which is independent of the number of carriers, given by $\kf$. Taking it
out, and neglecting logarithmic corrections, we find:
\begin{equation}
\delta \omega_{\vQ}^{{\rm bilayer}}  \propto  - \left( \frac{\partial
    t_0}{\partial l} \right)^2  
\frac{n}{M_C \omega_{\vQ} t_\perp} \, .
\label{shift_bilayer}
\end{equation}
We expect a similar dependence for other multilayer systems, as the main
ingredient in this estimate, the changes in the low energy density of
states in the two sublattices in each graphene layer, is independent of the
number of layers in the stack. If the carrier density differs significantly
among the layers\cite{G06}, we expect that phonons at each layer will be shifted by a
quantity which depends on the local charge. Note that a crossover to a shift typical of
single layer graphene will take place at $\omega_{\vQ} \sim \tp$. A similar
crossover will occur if $\mu \sim \tp$.

A bilayer system can show a gap in the electronic spectrum, when an applied
field or chemical doping breaks the symmetry between the two layers. In this case, the
electronic states close to the gap are mainly localized in one of the
layers. The polarizability shown in the diagram in Fig.~[\ref{phonon_ren}]
acquires a layer index, and is different in the two
layers. Hence, we expect that the in-plane phonons in each layer experience a
different frequency shift. In a first approximation, the phonons in the layer
where the states at the Fermi energy have highest weight show the
largest shift. Using dimensional arguments similar to those leading to
eq.~(\ref{susc_bilayer}), we expect that, when the Fermi wavevector is much
smaller than the wavevector at the center of the band of the biased bilayer,
$\kf \ll k_0 \approx \Delta / \vf$, the factor $n \sim \kf^2$ in
eq.(\ref{shift_bilayer}) is replaced by $k_0^2 \propto \Delta^2$.

The full dynamical response of a bilayer under a perpendicular applied field
is quite complex when the chemical potential is close to the gap
edges\cite{SPGN06}, with an anomalously large imaginary part. Hence, low
energy phonons in a biased bilayer should be significantly damped.

\section{Conclusions}
\label{conclusions}

We have analyzed the effect of a finite concentration of carriers on the
frequency shift of phonons in electrically doped graphene samples. We have
not considered changes due to modifications of the force constants,
associated to distortions of the $\sigma$ bonds. The analysis presented here 
shows that the shift in the optical phonon frequencies in electrically doped 
graphene samples can be observed in Raman experiments 
\cite{ferrari_06,eklund_06,ensslin_06,pinczuk}, 
and it can be used  to estimate the 
carrier density, or alternatively, the strength of the electron-phonon 
coupling. Notice that in an ordinary 2D electron gas the density of states 
(and charge susceptibility) depends only on the  electronic effective mass 
$m^*$ and is {\it independent} of the electronic density. Therefore, for an 
ordinary 2D electron gas the frequency shift is essentially uniform and 
independent of disorder. For Dirac fermions, however, because of the
effective Lorentz invariance in the continuum limit, we can write an 
equivalent of Einstein's relation between energy and mass: $\mu = m^* \vf^2$ 
(where the Fermi-Dirac velocity now plays the role of speed of light), 
indicating that the effective mass is energy dependent and vanishes
at the Dirac point ($\mu=0$). Therefore, the effect described here does not 
work in an ordinary 2D electron gas.  

Another interesting consequence of equation (\ref{shift}) is that the Raman 
shift should be larger close to extended effects such as edges, dislocations 
and cracks \cite{ferrari_06,eklund_06,ensslin_06}. The reason for that is the 
so-called {\it self-doping effect} discussed in great detail in 
ref.~[\onlinecite{nuno2006_long}]: because of the poor screening properties 
of Dirac fermions, the Coulomb interactions remain long-ranged and an
electrostatic potential builds up at the edges of the system, shifting the 
position of the surface states, and reducing the charge transfer to or from 
them. In this case the system, in order to maintain charge neutrality, can 
transfer charge to/from extended defects. This charging transfer is 
only halted when the charging energy of the edges is compensated by the
kinetic energy of the electrons. Thus, extra charge and a large density of 
states can be found at the edges of samples. In this case, according to 
eq.~(\ref{shift}), the Raman shift should change as a function of the distance 
from the sample edges (being larger at the edge). We have estimated that for
edges of size $0.1$-$1$ $\mu$ m the charge transfer is order of 
$10^{-4}$-$10^{-5}$ electrons per carbon ($\delta n \approx 10^{11} -10^{12}$
electrons per cm$^{-2}$) and hence the Raman shift is also of the order of 
a few wavenumbers but slightly larger than the effect produced by bulk
disorder. We also notice that this effect is not possible in the ordinary 2D 
electron gas because screening leads to a uniform charge distribution.

The shift of phonons with energies comparable or larger than the Fermi energy
is determined by the dynamic electronic response function, which is
significantly different from the static one. In this regime, the shift 
changes when $\mu \approx \omega_{\vQ}/2$, and vanishes at 
$\omega_\vQ \gg \mu$. 

Our results also suggest that the shift in phonon frequencies has a 
different dependence on carrier density in single layer and many layer systems,
eqs.~(\ref{shift}) and (\ref{shift_bilayer}). For a given carrier density, the
shifts in phonon frequencies should scale as $\delta \omega_{\vQ}^{1L}
\approx \delta \omega_{\vQ}^{2L} t_0 \sqrt{n_\Omega} / t_\perp$. Assuming
that $t_\perp / t_0 \approx 0.1$, the shift in a bilayer should be
smaller than in a single layer sample with the same carrier
concentration. This is consistent with experimental results which show that
the phonon frequencies in single layer systems are consistently lower than in
samples with many layers \cite{ferrari_06,eklund_06,ensslin_06}. 
The difference between a single layer and many layer systems is due to the 
fact that the low energy electronic wavefunctions has a reduced weight on the 
sites connected to other layers. Hence, it depends on the stacking order, and 
the shift is  different in samples with regions with rhombohedral structure, 
$(123123 \cdots) $\cite{GNP06}. In systems where the charge distribution
among the layers is not uniform, we expect that the shift of the phonons in 
different layers is also different, leading to a splitting of the single layer
phonon frequencies. Similar effects may also occur in bulk graphite 
\cite{Lanzara_annals}.
   
In summary, we have studied the effect of electronic inhomogeneities in the 
phonon spectrum of Raman active modes in graphene. We have shown that the 
electron-phonon coupling leads to a shift of the optical phonon frequency 
that is dependent on local electronic density. We argue that the frequency 
shift is larger at the edges than in the bulk of graphene and its value is 
of order of a few wavenumbers. These results have their origin on the 
Dirac-like nature of the quasi-particles in these materials and hence
do not have an analogue in the ordinary 2D electron gas. 

\section{Acknowledgments.}
We thank N.~M.~R.~Peres for many level-headed comments.
Discussions with K.~Ensslin, A.~Ferrari, A.~Geim, B.~Goldberg, D.~Graf,
Philip Kim, F.~Mauri, A.~Pinczuk, and A.~Swan, are acknowledged. 
A.H.C.N. is supported through NSF grant DMR-0343790.
F.G. acknowledges funding from MEC (Spain) through grant FIS2005-05478-C02-01
and the European Union Contract 12881 (NEST).
.
\appendix
\section{Analytical model for the in plane phonons in graphene.}
\label{anamodel}

\subsection{The model.}
The simplest model for the phonons of a single graphene plane includes only
nearest neighbor central forces\cite{G81}, following similar models for the
diamond lattice\cite{WA72}:
\begin{equation}
{\cal H} = \sum_{m,n} \frac{p_{mn}^2}{2 M} + \sum_{k,l;m,n} \frac{M
  \omega_0^2 [ ( {\bf a}_{kl} - {\bf a}_{mn} ) ( {\bf r}_{kl} - {\bf r}_{mn} )}{2}
\label{hamil}
\end{equation}
where the indices $k,l$ and $m,n$ label lattice sites which are nearest
neighbors.
The three possible orientations of the bonds attached to a given site, $m,n$,
allows us to define three unit vectors, ${\bf b}_{mn}^i , i=1,2,3$. We
define the displacement of the atom at site $m,n$, ${\bf r}_{mn}$  
by the three projections
$x_{mn}^i = {\bf b}_{mn}^i {\bf r}_{mn}$. These numbers satisfy
$\sum_{i=1,2,3} x_{mn}^i = 0$. The model contains a single parameter,
$\omega_0 = \sqrt{K/M}$, where $K$ is the spring constant of the bonds, and
$M$ is the mass of the Carbon atom.

\begin{figure}
\begin{center}
\includegraphics*[width=6cm]{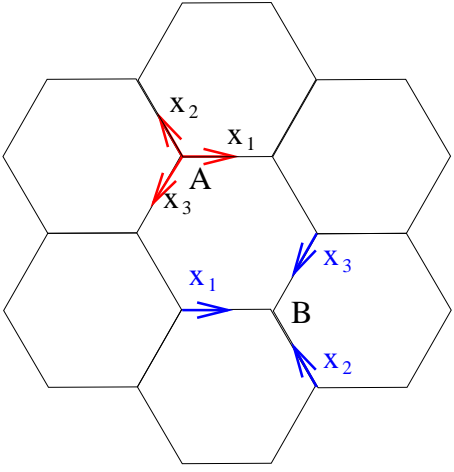}
\end{center}
\caption{Notation used for the atomic displacements used in the text.}
\label{lattice_phonons}
\end{figure}

\begin{figure}
\begin{center}
\includegraphics*[width=5cm,angle=-90]{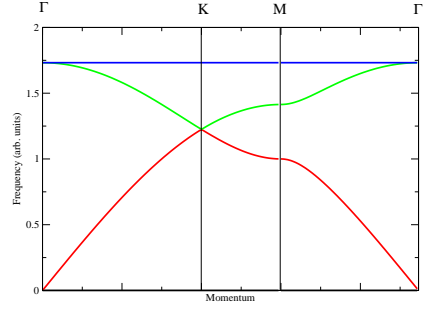}
\end{center}
\caption{Schematic dispersion relation of the phonons in the model used in
  the text.}
\label{dispersion_phonons}
\end{figure}

The equations of motion are:
\begin{equation}
\omega^2 x_{mn}^i = \omega_0^2 \left[ ( x_{mn}^i - x_{m'n'}^i ) - \frac{1}{2}
\sum_{j \ne i} ( x_{mn}^j - x_{m'' n''}^j ) \right]
\label{motion}
\end{equation} 
where the indices $m' n'$ and $m'' n''$ label sites which are nearest
neighbors to site $m n$ (see Fig.[\ref{lattice_phonons}]).
We now define the variable:
\begin{equation}
b_{mn} = x_{m' n'}^1 + x_{m'' n''}^2 + x_{m''' n'''}^3
\label{variable}
\end{equation}
using the displacements at the three sites connected to site $m n$ (see
Fig.[\ref{dispersion_phonons}]). In terms of these variables, the equations of
motion, eq.(\ref{motion}), can be written as:
\begin{equation}
\omega^2 b_{mn} = \frac{3}{2} \omega_0^2 b_{mn} + \frac{1}{2} \omega_0^2
\sum_{m' n'} b_{m' n'}
\label{motion_2}
\end{equation}
where $m' n'$ label the three sites connected to site $m n$. The calculation
of the phonon eigenstates is reduced to a tight binding model with one
orbital per site in the honeycomb lattice. From
eq.(\ref{motion_2}), we obtain two bands:
\begin{equation}
\frac{\omega_{\bf k}}{\omega_0}\!=\!
\sqrt{\frac{3}{2}\!\pm\!\frac{1}{2}\!\sqrt{3\!+\!2[\cos ( {\bf k} \cdot {\bf a}_1 )\!+\!\cos ( {\bf k} \cdot {\bf a}_2 )\!+\! \cos ( {\bf k} \cdot {\bf a}_3 )]}}
\label{dispersion}
\end{equation} 
where ${\bf a}_1 , {\bf a}_2$ are the unit vectors of the
honeycomb lattice, and and ${\bf a}_3 = {\bf a}_1 - {\bf a}_2$.  
From the knowledge of the variables $b_{mn}$ the equations of motion,
eq.(\ref{motion}), can be written as:
\begin{eqnarray}
\omega^2 x_{kl}^1 &= &\frac{3 \omega^2_0}{2} ( x_{kl}^1 - x_{mn}^1 ) -
\frac{\omega_0^2}{2} b_{mn} \nonumber \\ \omega^2 x_{mn}^1 &= &\frac{3
  \omega^2_0}{2} ( x_{mn}^1 - x_{kl}^1 ) - \frac{\omega_0^2}{2} b_{kl}
\label{displacement}
\end{eqnarray} 
From these equations the atomic displacements can be deduced from the set $\{
b_{mn} \}$.

The equations of motion, eq.(\ref{motion}) assume that all the variables
$b_{mn}$ are different from zero. When $b_{mn} = 0$, the equations
(\ref{displacement}) admit two additional solutions for $\omega^2 = 0$, 
and $\omega^2 = 3 \omega_0^2 / 2$. 
The bands obtained in eq.(\ref{dispersion}) correspond to
the longitudinal acoustical (LA) and longitudinal optical modes (LO). The two
additional flat bands obtained when $b_{mn}=0$ describe the transverse
acoustical (TA) and transverse optical (TO) modes. The phonon bands are shown
in Fig.[\ref{dispersion_phonons}]. The existence of a flat TA band at $\omega
= 0$ reflects the band that the honeycomb lattice can be distorted without
changing the distance between nodes. These deformations do not have an energy
cost in a nearest neighbor central forces model described in
eq.(\ref{hamil}). The velocity of sound of the LO modes is $v_{\rm s} = (
\omega_0 a ) / ( 2 \sqrt{2} )$, where $a$ is the lattice constant.

\subsection{Defects}

The mapping to a scalar tight binding model of the Hamiltonian in
eq.(\ref{hamil}) can be extended to lattices with defects. We describe the
defect as the absence of bonds. Hence, an atom near a defect is attached to
fewer neighbors than one at the bulk. This implies that the condition $\sum
x_{mn}^i = 0$ is no longer satisfied. We can take this into account by
defining a new variable at the sites near the defect, $a_{mn} = \sum_i'
x_{mn}^i$, where the sum is restricted to the bonds which remain intact. 

\subsubsection{Zigzag edge}

\begin{figure}
\begin{center}
\includegraphics*[width=6cm]{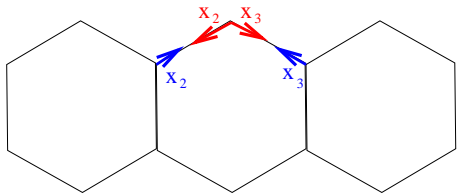}
\end{center}
\caption{Displacements used to define the variable $a_{mn}$ at an atom at a zigzag edge
(red), and the variable $b_{mn}$ (blue).}
\label{edge_phonon}
\end{figure}

The atoms at a zigzag edge are connected by only two bonds to the rest of the
lattice. We define the variable $b_{mn}$ using the displacements of the two nearest
neighbor atoms to the edge atom $mn$. The atomic displacements used to define
the variables $a_{mn}$ and $b_{mn}$ are sketched in Fig.[\ref{edge_phonon}].
The equations of motion for the
variables $a_{mn}$ and $b_{mn}$ when
the indices $mn$ label an atom at the edge become:
\begin{eqnarray}
\omega^2 a_{mn} &= &\frac{\omega_0^2}{2} ( a_{mn} - b_{mn} ) \nonumber \\
\omega^2 b_{mn} &= &\frac{3 \omega_0^2}{2} b_{mn} - \frac{\omega_0^2}{2}
\sum_{m' n'} b_{m' n'} - \frac{\omega_0^2}{2} a_{mn} \nonumber \\
\omega^2 b_{m'n'} &= &\frac{3 \omega_0^2}{2} b_{m'n'} - \frac{\omega_0^2}{2}
\sum_{m'' n''} b_{m'' n''} - \frac{\omega_0^2}{2} a_{mn}
\label{edge}
\end{eqnarray}
where the indices $m'n'$ label the sites which are the nearest neighbors of
the vacancy, and $m'' n''$ stand for the next nearest neighbors. 
 The equations of motion for the remaining atoms are not changed from
eq.(\ref{motion_2}). 
\begin{figure}
\begin{center}
\includegraphics*[width=7cm,angle=-90]{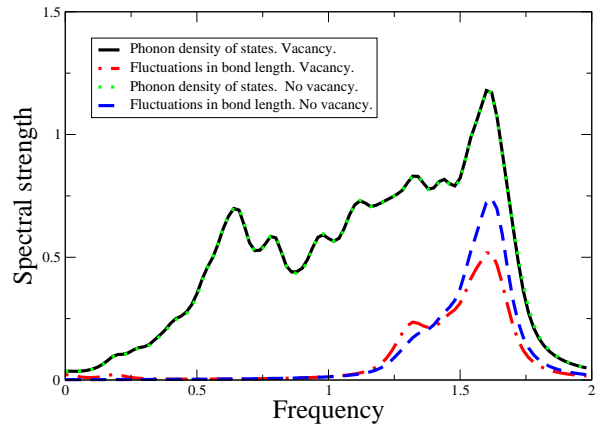}
\end{center}
\caption{Phonon density of states and spectral function of bond fluctuations
  with and without a vacancy.} \label{phonon_dos}
\end{figure}

Thus, the equations of motion of the atoms can be mapped onto a tight binding
model. The only difference with the bulk case is that the description of the
displacements of the atoms at the boundary require the definition of two
effective orbitals. The position of the effective orbital level $a_{mn}$ 
 at the edge, $\omega_0^2 / 2$, is lower than that for the variable $b_{mn} ,
 (3 \omega_0^2 ) / 2$.  This reflects the fact that atomic fluctuations
 are enhanced at  the edge.

\subsubsection{Vacancy.}

As in the case of an atom at a zigzag edge, the three atoms near a vacancy
are connected by bonds to two nearest neighbors only. As in the previous
case, a new variable, $a_{mn} = \sum_i' x_{mn}^i$ needs to be defined at
these three sites. The equations of motion for the variables $a_{mn}$ and
$b_{mn}$ are those in eq.(\ref{edge}).

The phonon density of states in clusters with and without vacancies, and the
spectral strength of the bond length fluctuations, in the bulk and near a
vacancy are shown in Fig.[\ref{phonon_dos}]. Contrary to what happens for the
$\pi$ electronic band, the phonons are not too disturbed near a vacancy,
although some shift of spectral strength to lower energies takes place.

\bibliography{raman_bib}
\end{document}